\documentstyle[prl,aps,epsf]{revtex}

\begin{document}

\twocolumn[\hsize\textwidth\columnwidth\hsize\csname@twocolumnfalse\endcsname
\begin{center}
{\large
On the gauge invariance of the DVCS amplitude }

\it{
I.V.~Anikin$^{a}$, B.~Pire$^{b}$, O.V.~Teryaev$^{ab}$}

\address{
{\it $^{a}$ Bogoliubov Laboratory of Theoretical Physics,
JINR, Russia} \\
{\it $^{b}$ CPhT,
{\'E}cole Polytechnique,
91128-Palaiseau, France } }
\date{\today}
\maketitle

\end{center}

\begin{abstract}

We analyze in detail the problem of gauge invariance of the deeply
vitual Compton scattering (DVCS) amplitude. Using twist 3 one-gluon
exchange diagram contributions and the QCD equations of motion,
we derive the general gauge invariant expression of the DVCS amplitude
on a (pseudo)scalar particle (pion, $He^4$).
Similarly to the case of deep inelastic scattering, the amplitude does
not depend on the twist-3 quark-gluon correlations at the Born level.
The contribution of the derived amplitude to the single-spin asymmetry
with longitudinally polarized lepton is calculated.
\end{abstract}
\vspace{1cm}
]

Deeply Virtual Compton Scattering (DVCS) has recently attracted
much attention. One of the main reasons of this interest is the fact
that the DVCS process gives information about a new type of
parton distributions, called skewed  parton distribution
(see for example \cite{DM,Ji,Rad,Gui98} and references therein).
The process

\begin{equation}
\label{pro}
\gamma^*(q) N (p) \to \gamma(q') N (p')
\end{equation}

\noindent
has been shown to factorize in the Bjorken region with
$(q^{\prime})^2=0$, $-q^2$
large and small transfer $t = (p-p')^2$, as the product of a
perturbatively calculable coefficient function and a long distance
object, the skewed  parton distribution, which generalizes the notion of
parton distributions.

The fact that there is a problem with the
photon gauge invariance of the DVCS amplitude in leading order at Bjorken
limit is fairly well-known (see, for instance \cite{Gui98}).
The relevant terms are proportional to the transverse component of the
momentum transfer and provide the leading contribution to some
observables, and in particular, to the Single Spin Asymmetry.

As was shown in \cite{Pire}, a fruitful analogy
between the transverse spin case of the deep inelastic scattering (DIS)
and the DVCS process can be used to derive a general solution of this
problem. We elaborate on this approach in the current paper.

For simplicity, we concentrate here on the DVCS
process off (pseudo)scalar hadrons, which may be pions or helium-4
nuclei, but  our calculation may be generalized to any hadrons.
We also neglect  hadron masses effects, i.e.
kinematical power corrections, which may be studied independently.

The Lorentz structure of the hard subgraph of
the leading order DVCS diagram (Fig.1a)
has the  form of a transverse projector
$g^{\mu \nu}-P^{\mu} n^{\nu}-n^{\mu} P^{\nu}$ \cite{Ji}.
All $4$-vectors  may be presented in the
form of the Sudakov decomposition over two light-cone vectors $P,n$
and one component transversal to the given light-cone vectors.
Hence, if the virtual photon momentum, which has a transverse
component, is convoluted
with the hard part of the leading order DVCS amplitude,
one obtains a term directly proportional to the transverse component
of the virtual photon momentum. In other words, the measure of
the photon gauge invariance violation is
the non-zero transverse component of the virtual photon momentum.
The recent general analysis \cite{BR} confirmed that violating terms
are indeed kinematically subleading.

In this paper we generalize
the Ellis-Furmanski-Petronzio (EFP) factorization scheme \cite{E&F&P} to
the non-forward case and calculate the complete expression for the
DVCS amplitude up to twist 3 order.
While this scheme  was originally applied to the case of the twist-4 power
corrections to DIS, our analysis will be more close to the
subsequent treatment of the transverse polarization in DIS at twist 3 level
\cite{Efr84,A&E&L},
Analogously to that case, the process (\ref{pro})
is described by the diagrams of Fig.1, which contain leading and next
to leading twist contributions.
Namely, the diagram (b) , in the case of the transverse gluon field, is
entirely at twist 3 level, while the handbag diagram (a)
contains, besides the standard twist-2 term (produced by the good
component of quark fields and collinear parton momenta), a twist-3 term,
related to the quark gluon contribution of diagram (b) by the equations
of motion. The latter play a crucial role in DIS, guaranteeing that
the sum of diagrams (a) and (b) is gauge invariant and, moreover,
not explicitely dependent of the quark-gluon correlations.
We will show that the situation in DVCS is quite similar.

The sum of $T_{\mu\nu}^{(a)}$ amplitude from diagram (a)
and $T_{\mu\nu}^{(b)}$ from diagram (b),
has the following form, like in \cite{Efr84},
\begin{eqnarray}
\label{1}
&&T_{\mu\nu}^{(a)}+T_{\mu\nu}^{(b)}=
\int dk {\rm tr} \biggl\{ E_{\mu\nu}(k) \Gamma (k) \biggr\}+
\nonumber\\
&&\int d k_1 d k_2{\rm tr}\biggl\{
E_{\mu\rho\nu}(k_1, k_2) \Gamma_{\rho} (k_1, k_2) \biggr\}
\end{eqnarray}
where $E_{\mu\nu}$ and $E_{\mu\rho\nu}$ are the coefficient functions
with two quark legs and two quark and one gluon legs,
respectively. For simplicity,
we restrict to the Born diagrams for the coefficient functions.
In (\ref{1}), the following notations are introduced
\begin{eqnarray}
\label{2}
&&\Gamma_{\alpha\beta}(k)=-\int dz e^{i( k-\frac{\Delta}{2}) z }
\langle p^{\prime}|
\psi_{\alpha}(z) \bar\psi_{\beta}(0)| p \rangle ,
\nonumber\\
&&\Gamma^{\rho}_{\alpha\beta} (k_1, k_2)=
-\int d z_1 d z_2
e^{ i( k_1-\frac{\Delta}{2}) z_1 +i(k_2-k_1) z_2 }
\nonumber\\
&&\langle p^{\prime} |
\psi_{\alpha}(z_1) g A^{\rho}(z_2)
\bar\psi_{\beta}(0)| p \rangle ,
\nonumber\\
&&p^{\prime}=P+\frac{\Delta}{2}, \quad
p=P-\frac{\Delta}{2}, \quad \Delta= q-q^{\prime}
\end{eqnarray}
Here, $p^{\prime}$ and $p$ are the final and initial hadron momenta,
$q^{\prime}$ and $q$ are the final and initial photon momenta.
For the sake of convenience, we neglect all the kinematical power
corrections and put $P^2$ and $t\equiv\Delta^2$ both equal to zero,
keeping only the terms linear in $\Delta_T$.
Also, we choose the axial gauge condition for
gluons, {\it i.e.} $n\cdot A =0$, where $n$ is a light-cone vector,
normalized by the condition $n\cdot P =1$.
It is convenient to assume that $n=q^{\prime}/P\cdot q^{\prime}$,
although our result will not depend explicitely on the choice of $n$.
We carry out a
decomposition of  $k$ in the basis defined by
 the $P$-- and $n$-- light-cone
vectors
\begin{eqnarray}
\label{k}
k=xP + (k\cdot P)n + k_T,
\quad x=k\cdot n .
\end{eqnarray}
Apart from this,
\begin{eqnarray}
\label{del}
\Delta=-2\xi P + \Delta_T.
\quad -2\xi=\Delta\cdot n.
\end{eqnarray}
Further, we carry out the following replacement for the integration
momentum in eqn. (\ref{1})
\begin{eqnarray}
\label{subst}
dk_i\, \to\, dk_i dx_i \delta(x_i-k_i\cdot n).
\end{eqnarray}
Expanding the two-quark coefficient function $E_{\mu\nu}$ (see, (\ref{1}))
in a Taylor series
and, next, using the following Ward identity \cite{E&F&P,Efr84}
\begin{equation}
\frac{\partial E_{\mu\nu}(k)}{\partial k^\rho}=E_{\mu\rho\nu}(k,k)
\end{equation}

we can write the DVCS amplitude as
\begin{eqnarray}
\label{1.1}
&&T_{\mu\nu}^{(a)}+T_{\mu\nu}^{(b)}=
\int dx{\rm tr}\biggl\{ E_{\mu\nu}(x P) \Gamma (x) \biggr\}
+
\nonumber\\
&&\int dx_1 dx_2{\rm tr}\biggl\{
E_{\mu\rho\nu}(x_1 P, x_2 P) \omega_{\rho\rho^{\prime}}
\Gamma_{\rho^{\prime}} (x_1, x_2) \biggr\}
\end{eqnarray}
where $\omega_{\rho\rho^{\prime}}=\delta_{\rho\rho^{\prime}}-
n_{\rho^{\prime}} P_{\rho}$,
and
\begin{eqnarray}
\label{2.1}
&&\Gamma_{\alpha\beta}(x)=-\int d\lambda \,
e^{ i( x+\xi)\lambda }
\langle p^{\prime} |
\psi_{\alpha}(\lambda n) \bar\psi_{\beta}(0)| p \rangle ,
\nonumber\\
&&\Gamma^{\rho^{\prime}}_{\alpha\beta} (x_1, x_2)=
-\int d\lambda_1 d\lambda_2 \,
e^{ i( x_1+\xi) \lambda_1
+i(x_2-x_1) \lambda_2 }
\nonumber\\
&&\langle p^{\prime} |
\psi_{\alpha}(\lambda_1 n) D^{\rho^{\prime}}(\lambda_2 n)
\bar\psi_{\beta}(0)| p \rangle ,
\end{eqnarray}
where $D_{\mu}$ is the QCD covariant derivative in the fundamental
representation.

Let us now focus on the
QCD equations of motion both for
incoming and outcoming quarks.
{}From these equations,
we deduce integral relations for structure functions,
parametrizing quark and quark-gluon correlations.
So, let us to start from the QCD equations of motion
(we consider massless quarks)
\begin{eqnarray}
\langle
\overrightarrow{
\hat D(z)} \psi(z) \bar\psi(0)\rangle = 0
\quad
\langle\psi(z) \bar\psi(0)
\overleftarrow{\hat D(0)} \rangle  = 0,
\end{eqnarray}
where $\langle\ldots\rangle$ denote the asymmetrical matrix elements.
Keeping only vector and axial projections (since for massless quarks
all other structures do not contribute),
we decompose the
quark and quark-gluon correlators in the $\gamma$-basis.
We have
\begin{eqnarray}
\label{qcor}
&&-4\langle \psi(z) \bar\psi(0) \rangle =
\langle \bar\psi(0) \gamma_{\alpha} \psi(z) \rangle \gamma_{\alpha}-
\langle \bar\psi(0) \gamma_{\alpha}\gamma_5
\psi(z) \rangle \gamma_{\alpha}\gamma_5
\nonumber\\
\\
\label{qgcor}
&&-4\langle g A^{\rho}(y)\psi(z) \bar\psi(0) \rangle =
\langle \bar\psi(0) \gamma_{\alpha} g A^{\rho}(y)\psi(z)
\rangle \gamma_{\alpha}-
\nonumber\\
&&
\langle \bar\psi(0) \gamma_{\alpha}\gamma_5
g A^{\rho}(y)\psi(z) \rangle \gamma_{\alpha}\gamma_5.
\end{eqnarray}
In terms of $P, \Delta^{T}, n$- vectors,
we introduce
the parametrization of relevant vector and axial correlators
(see, Eq. (\ref{qcor}, \ref{qgcor}))
in the following forms, where terms proportional to $n$,
contributing only at twist-4 level, have been omitted,
and the axial gauge condition $n\cdot A =0$ has been taken into account~:
\begin{eqnarray}
\label{par1}
&& \langle \bar\psi(0)\gamma_{\mu} \psi(z) \rangle
\stackrel{{\cal F}}{=}
H_1(x)P_{\mu}+H_3(x)\Delta^{T}_{\mu},
\\
\label{par11}
&& \langle \bar\psi(0)\gamma_{\mu}\partial^T_{\rho} \psi(z) \rangle
\stackrel{{\cal F}}{=}
i H_1^T(x)P_{\mu}\Delta^T_{\rho},
\\
\label{par1.1}
&& \langle \bar\psi(0)\gamma_5\gamma_{\mu} \psi(z) \rangle
\stackrel{{\cal F}}{=}
iH_A(x)\varepsilon_{\mu\Delta^{T}Pn},
\\
\label{par1.11}
&& \langle \bar\psi(0)\gamma_5\gamma_{\mu}
\partial^T_{\rho} \psi(z) \rangle
\stackrel{{\cal F}}{=}
-H_A^T (x) P_{\mu}\varepsilon_{\rho\Delta^{T}Pn},
\\
\label{par1.2}
&& \langle \bar\psi(0)\gamma_{\mu}g A_{\rho}^T(y) \psi(z) \rangle
\stackrel{{\cal F}}{=}
B(x_1,x_2)P_{\mu}\Delta^{T}_{\rho},
\\
\label{par1.3}
&& \langle \bar\psi(0)\gamma_5\gamma_{\mu} g A_{\rho}^T(y) \psi(z) \rangle
\stackrel{{\cal F}}{=}
i D(x_1,x_2)P_{\mu}\varepsilon_{\rho\Delta^{T}Pn},
\end{eqnarray}
here $
\varepsilon_{\rho\Delta^{T}Pn} \equiv \varepsilon_{\rho \alpha \beta \gamma}
\Delta^{T \alpha} P^\beta n^\gamma~;
\stackrel{{\cal F}}{=}$ denotes the Fourier transformation
with measure ($z=\lambda n, z^{\prime}=0 $)
$$
dx e^{ -i( xP-\frac{\Delta}{2})z
+i( xP+\frac{\Delta}{2})z^{\prime} }
$$
for quark correlators, and
$$
dx_1 dx_2 e^{ -i( x_1P-\frac{\Delta}{2})z
-i(x_2 - x_1)Py
+i( x_2P+\frac{\Delta}{2})z^{\prime} }
$$
for quark-gluon correlators. Note that the latter for the non-forward
case are actually new objects.
We kept only the  argument $x$ for all the correlators,
dropping for brevity the dependence of the distributions on the
skewedness parameter $\xi$, recovering it below
where it is necessary. Their dependence on
$t=\Delta^2$
is beyond our scope.
Finally, their dependence on the factorization scale parameter
$\mu^2$
requires an extended separate investigation.

The dependence on $\xi$ plays a crucial role in the
disappearance of the $H_3$ term, when a local current,
related to its integral in $x$, is considered. Although
this is required by the conservation of local vector current,
this effect is proportional to $t$ and is therefore beyond the scope of our
approximation. At the same time, this is also required by
T-invariance \cite{PW}. It is therefore natural,
that symmetry of $H$ (c.f. \cite{man,Gui98}),
resulting also from T-invariance,
$$
H_3(x,\xi)=-H_3(x,-\xi)
$$
is relevant. Calculating the integral in $x $ and implying the
polynomiality condition \cite{Ji}, one gets a function which
is independent on $\xi$, and hence vanishes since the only odd
constant is zero. A similar argument is also applicable to the
function $H_A$, so that~:

\begin{eqnarray}
\label{n-ind5}
\int dx H_3(x) = 0~,  \qquad
\int dx H_A(x) = 0.
\end{eqnarray}

Acting $i\hat\partial$ on (\ref{qcor}) and  $\gamma$ on (\ref{qgcor})
from left or right side, the QCD equations of motion yield the
following integral relations for structure functions
\begin{eqnarray}
\label{me2.1.1}
&&\int dy \left( B^{(A)}(x,y)-D^{(S)}(x,y)
+ \delta(x-y)H_A^T(y) \right) =
\nonumber\\
&&-\xi H_3(x) - \frac{1}{2}H_1(x) - x H_A(x) ,
\nonumber\\
&&\int dy \left(B^{(S)}(x,y) - \delta(x-y)H_1^T(y)
-D^{(A)}(x,y)\right) =
\nonumber\\
&&xH_3(x)+\xi H_A(x),
\end{eqnarray}
where symmetrical and anti-symmetrical functions are defined as,
\begin{eqnarray}
\label{saf}
&& B^{(S,A)}(x,y)=\frac{1}{2}
\left( B(x,y) \pm B(y,x) \right).
\end{eqnarray}

Note again the important difference with DIS, where the axial
correlator is symmetric and the vector one is antisymmetric \cite{Efr84}.
The latter property is based on  T-invariance, just like the
symmetry properties in $\xi$ discussed above. To see the relation
between $x \leftrightarrow y$ and $\xi$ symmetry, it is instructive
to write the general T-invariance relations~:
\begin{eqnarray}
\label{TG}
B(x,y,\xi) = B (y,x,-\xi), \,\,
D(x,y,\xi) = - D (y,x,-\xi);
\end{eqnarray}
So the "unnatural" symmetry in $x,y$ results from the antisymmetrical
in $\xi$ part, clearly absent in the forward case. It is worthy to
note that a similar unnatural symmetry may appear due to the
final state interaction phases in the case of T-odd fragmentation
functions \cite{OT96}.

We can thus write the DVCS amplitude in a
gauge invariant manner. Let us first write the contribution from
the pure quark amplitude~:
\begin{eqnarray}
\label{qT}
&&T_{\mu\nu}^{(a)} = \int dx\frac{1}{(xP+Q)^2}
\nonumber\\
&&\Biggl(
H_1(x) S_{\nu (xP+Q)\mu P} + H_3(x)S_{\nu (xP+Q)\mu\Delta^T} +
\nonumber\\
&&H_A(x)\varepsilon_{\alpha\Delta^T Pn}
\varepsilon_{\nu (xP+Q)\mu\alpha}
\Biggr) + \left( \mu\to\nu, \, Q\to -Q \right),
\end{eqnarray}
where the following notations were introduced
\begin{eqnarray}
&&S_{\mu_1\mu_2\mu_3\mu_4}=g_{\mu_1\mu_2}g_{\mu_3\mu_4}+
g_{\mu_1\mu_4}g_{\mu_2\mu_3}-g_{\mu_1\mu_3}g_{\mu_2\mu_4},
\nonumber\\
&& Q=(q+q^{\prime})/2
\nonumber
\end{eqnarray}
As a corollary,
the contribution of the amplitude corresponding to the one-gluon exchange
diagram, has the form
\begin{eqnarray}
\label{qgT}
&&T_{\mu\nu}^{(b)} =
\frac{1}{4}\int dx_1dx_2 \frac{1}{(x_1P+Q)^2(x_2P+Q)^2}
\nonumber\\
&&\Biggl(
\biggl( B(x_1,x_2)-\delta(x_1-x_2)H_1^T(x_2)\biggr)
\Biggr.
\nonumber\\
\Biggl.
&&{\rm tr}\biggl(
\gamma_{\nu}(x_2\hat P+\hat Q)\hat\Delta^T(x_1\hat P+\hat Q)
\gamma_{\mu}\hat P \biggr) +
\Biggr.
\nonumber\\
\Biggl.
&&i\biggl( D(x_1,x_2) - \delta(x_1-x_2)H_A^T(x_2) \biggr)
\varepsilon_{\alpha\Delta^T Pn}
\Biggr.
\nonumber\\
\Biggl.
&&{\rm tr}\biggl(
\gamma_{\nu}(x_2\hat P+\hat Q)\gamma_{\alpha}(x_1\hat P+\hat Q)
\gamma_{\mu}\hat P\gamma_5 \biggr)
\Biggr) + "crossed".
\end{eqnarray}
Calculating all traces in (\ref{qgT}),
using the following obvious identities
\begin{eqnarray}
&&\frac{\pm(P\cdot Q)(x_1+x_2)+ Q^2 }{(x_1P\pm Q)^2(x_2P\pm Q)^2}=
\nonumber\\
&&\frac{1}{2}
\Biggl(
\frac{1}{(x_1P\pm Q)^2}+\frac{1}{(x_2P\pm Q)^2}
\Biggr),
\nonumber\\
&&\frac{\pm (P\cdot Q)(x_1-x_2)}{(x_1P\pm Q)^2(x_2P\pm Q)^2}=
\nonumber\\
&&\frac{1}{2}
\Biggl(
\frac{1}{(x_2P\pm Q)^2}-\frac{1}{(x_1P\pm Q)^2}
\Biggr),
\nonumber
\end{eqnarray}
and the equations of motion (\ref{me2.1.1}) in terms
of symmetric and anti-symmetric functions,
we add contributions of (\ref{qT}) and
(\ref{qgT}), taking into account the crossed diagrams.
The gauge invariant expression of the DVCS amplitude
is thus
\begin{eqnarray}
\label{amp}
T_{\mu\nu} =
-\frac{1}{2P\cdot Q}\int dx
\Biggl(
\frac{1}{x-\xi+i\epsilon} + \frac{1}{x+\xi-i\epsilon}
\Biggr)
{\cal T}_{\mu\nu},
\nonumber\\
\end{eqnarray}
where
\begin{eqnarray}
&&{\cal T}_{\mu\nu} =
H_1(x)
\Biggl(
-2\xi P_{\mu}P_{\nu}  -
P_{\mu}Q_{\nu} - P_{\nu}Q_{\mu} +
\Biggr.
\nonumber\\
\Biggl.
&&g_{\mu\nu}(P\cdot Q) - \frac{1}{2}P_{\mu}\Delta_{\nu}^{T} +
\frac{1}{2}P_{\nu}\Delta_{\mu}^{T}
\Biggr) -
\nonumber\\
&&H_3(x)
\Biggl(
\xi P_{\nu}\Delta_{\mu}^T + 3\xi P_{\mu}\Delta_{\nu}^T
+ \Delta_{\mu}^{T}Q_{\nu} + \Delta_{\nu}^{T}Q_{\mu}
\Biggr) +
\nonumber\\
&&\frac{\xi}{x}H_A(x)
\Biggl(
3\xi P_{\mu}\Delta_{\nu}^{T} -
\xi P_{\nu}\Delta_{\mu}^{T} -
\Delta_{\mu}^{T}Q_{\nu} + \Delta_{\nu}^{T}Q_{\mu}
\Biggr).
\nonumber
\end{eqnarray}
We can see that the first term of (\ref{amp}), proportional to
the $H_1$-function,
completely coincides with the improved DVCS amplitude, proposed by Guichon
and Vanderhaegen (GV) in \cite{Gui98}.
Indeed, if we proceed, in such terms, to GV
basis \cite{Gui98}, where the $n$-vector
is expressed via the virtual photon momentum $q$
and decompose the $Q$-vector in this basis, then we derive, omitting
the terms with $H_3$- and $iH_A$- functions~:
\begin{eqnarray}
\label{amG}
&&T_{\mu\nu} =
\int dx\frac{1}{2}
\Biggl(
\frac{1}{x-\xi+i\epsilon
} + \frac{1}{x+\xi-i\epsilon}
\Biggr)\cdot
\nonumber\\
&&H_1(x)
\Biggl(
P_{\mu}n_{\nu} + P_{\nu}n_{\mu} - g_{\mu\nu}
- \frac{P_{\nu}\Delta_{\mu}^{T}}{P\cdot q}
\Biggr) \equiv
\nonumber\\
&& T^{L.O.}_{\mu\nu} + \frac{P_{\nu}}{P\cdot q}\Delta^T_{\lambda}
T^{L.O.}_{\mu\lambda},
\end{eqnarray}
where the definition of $T^{L.O.}_{\mu\nu}$ amplitude is related with
the transverse direction projector, which is
$(P_{\mu}n_{\nu} + P_{\nu}n_{\mu} - g_{\mu\nu})$.

In complete analogy to DIS, the answer does not depend explicitely
on quark-gluon correlations. However, in contrast to DIS, it contains
two additional new functions, instead of the single function
$g_2$ in the DIS case. While $H_3$ may be considered
as an analog of $g_2$ (the coefficients of them are $\Delta_T$ and
$s_T$, respectively), the DIS analog of the
function $H_A$ is excluded by
T-invariance and may be present only for fragmentation.

We emphasize that the deduced gauge
invariant expression of the DVCS amplitude has a significant meaning
for the sequential application of QCD to the investigation of any observable
values. To demonstrate this, let us consider the single (electron)
spin asymmetry (SSA), which arises in the collision of the longitudinaly
polarized electron beams with the unpolarized scalar target.
The SSA parameter
(for details see, for instance, \cite{Ji}) is
\begin{eqnarray}
\label{ssa}
{\cal A}_L=\frac{d\sigma(\rightarrow)-d\sigma(\leftarrow)}
{d\sigma(\rightarrow)+d\sigma(\leftarrow)},
\end{eqnarray}
where
\begin{eqnarray}
\label{dsec}
&&d\sigma(\rightarrow)-d\sigma(\leftarrow)\sim
\frac{e^6F_+(t) 2\xi}{q^2 t (k-\Delta)^2 (k^{\prime}+\Delta)^2}
\varepsilon_{kk^{\prime}P \Delta}
\nonumber\\
&&\int dx \Biggl( \delta(x+\xi)-\delta(x-\xi) \Biggr)\cdot
\Biggl(
H_1(x)(( k + k^{\prime})\cdot P)
+
\Biggr.
\nonumber\\
\Biggl.
&&2 H_3(x)\biggl(
\xi((k+k^{\prime})\cdot P) + (k^{\prime}\cdot\Delta) -
\xi(P\cdot Q)
\biggr) +
\Biggr.
\nonumber\\
\Biggl.
&&\frac{2\xi}{x(P\cdot Q)}H_A(x)\biggl(
(k\cdot\Delta)(k^{\prime}\cdot P) - (k^{\prime}\cdot\Delta)(k\cdot P)
\biggr)
\Biggr) +
\nonumber\\
&&
\frac{e^6}{q^4}\varepsilon_{kk^{\prime}P \Delta }
\frac{2\xi}{(P\cdot Q)}
\int dxdx^{\prime}
\nonumber\\
&&\Biggl(
\biggl[ \delta(x+\xi)-\delta(x-\xi) \biggr]
\biggl[ \frac{{\cal P}}{x^{\prime}-\xi}+
\frac{{\cal P}}{x^{\prime}+\xi} \biggr] -
\Biggr.
\nonumber\\
&&\Biggr.
\biggl[ \delta(x^{\prime}+\xi)-\delta(x^{\prime}-\xi) \biggr]
\biggl[ \frac{{\cal P}}{x-\xi}+\frac{{\cal P}}{x+\xi} \biggr]
\Biggr)\cdot
\nonumber\\
&&\Biggl(
H_1(x)H_3(x^{\prime}) - H_1(x^{\prime})H_3(x) +
\Biggr.
\nonumber\\
&&\Biggl.
\biggl[ H_1(x^{\prime})\frac{H_A(x)}{x} -
H_1(x)\frac{H_A(x^{\prime})}{x^{\prime}}
\biggr] \xi
\Biggr),
\end{eqnarray}
where $F_+(t)$ is the target electromagnetic form factor, emanating
from the Bethe-Heitler diagrams, $k$ and $k^{\prime}$ denote the
momenta of the initial and final electron $(k-k^{\prime}=q)$.

The authors would like to thank
A.P. Bakulev, A.E. Dorokhov, M. Diehl,
A.V. Efremov, E.A. Kuraev, N.I. Kochelev,
S.V. Mikhailov, D. M{\"u}ller,
A.V. Radyushkin, R. Ruskov and A. Sch{\"a}fer
for useful discussions and comments. CPhT is UMR 7644 du CNRS.

\begin{figure}
\vspace*{2.cm}
\hspace*{0.7cm}
\epsfxsize=12cm
\epsfysize=8cm
\centerline{\epsfbox{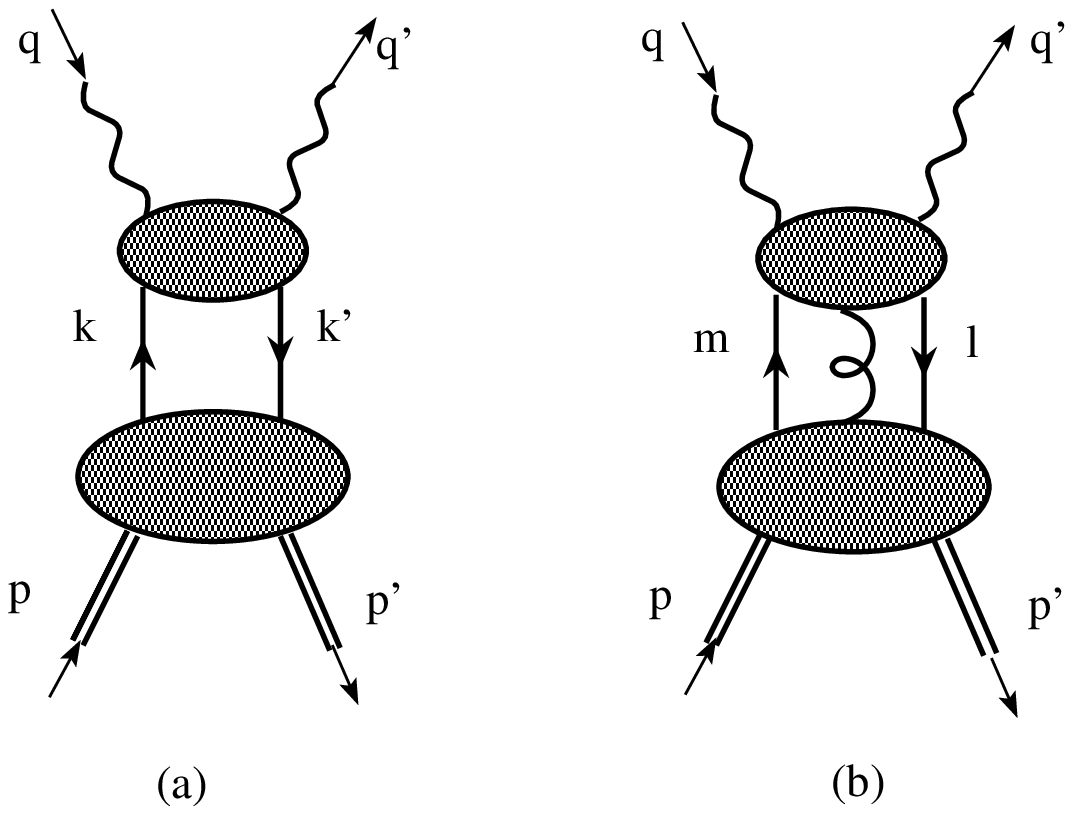}}
\end{figure}
\vspace*{-1.cm}
Fig.1 The DVCS diagrams ( notations:
$k=xP-\Delta/2$,  $k^{\prime}=xP+\Delta/2$,
$m=x_1P-\Delta/2$ and $l=x_2P+\Delta/2$).

\end{document}